\documentclass[twoside]{ilcws10}
\usepackage[latin1]{inputenc}
\usepackage[dvips]{graphicx,epsfig,color}
\usepackage{wrapfig,rotating}
\usepackage{amssymb,amsmath,array}

\begin{document}
\title{
Influence of beam related background on ILD reconstruction} 
\author{Katarzyna Wichmann 
\vspace{.3cm}\\
DESY \\
Notkestr. 85, 22607 Hamburg, Germany
}

\maketitle

\begin{abstract}
At the ILC machine induced backgrounds, mainly electron-positron pairs from beamstrahlung,
will be the most important source of unwanted interactions. 
Methods of generating, simulating and imposing pair-background on physics events are presented
and influence of this background on ILD reconstruction is reviewed.
Effects of the background on various ILD detectors are shown for nominal beam parameters and for
SB09 reduced beam parameter set 
and an example of a physics analysis performed with overlayed pair background is given.
\end{abstract}

\section{Beam induced backgrounds at the Linear Collider}

Lepton linear colliders face a novel problem - beam induced backgrounds. Effects of these backgrounds at the present
colliders are negligible and so far were not studied or taken into account. However at the ILC machine induced backgrounds
will be the most important source of unwanted interactions and therefore should be simulated, studied and taken into account
in physics analysis with great care. Detail description of the beam induced background can be found 
elsewhere~\cite{ref-adrian}, here only the most important component is shortly described.

The most prominent beam related background are so called ``beamstrahlung'' photons coming from beam-beam interactions and 
$e^+e^-$ pairs produced by these photons. The photons are strongly focused in the forward direction and exit through the
beam tube, being of no consequence for the physics events at the interaction point. The electron-positron pairs are 
a different case. They can easily reach the detector, both the direct particles and the scattered ones. 
Most affected parts of the detector are vertex detector and forward detectors. For ILC-like beam parameters 
there is on average about $10^5$ pairs per bunch crossing of total energy of $100$~TeV with an average energy per particle of about 
few GeV. 
That shows that the $e^+e^-$ pairs should be treated with care, especially that they are an unavoidable background. 
The machine and detector designs take into 
account possible ways of reducing the $e^+e^-$ pairs spray reaching the detector but the remaining background needs to be
taken into account on the reconstruction and analysis level.

\section{Simulating beam backgrounds}

The $e^+e^-$ pairs can be simulated using various generators:  Ginuea-Pig generator (GP)~\cite{ref-gineapig}, 
Ginuea-Pig++~\cite{ref-gp++} and CAIN~\cite{ref-cain}. Comparisons between GP and CAIN (they use the same models but
a different implementation) show good agreement on the level of 10\% between these generators. 
Guinea-Pig++ (GP++) is a newly developed object oriented version of the C code Guinea-Pig. It implements new features and code
extensions and the simulated variables are backward-compatible with the results from GP.
For the ILD simulation a full detector simulation Mokka~\cite{ref-mokka} has been used, which is a standard tool for ILD detector 
simulation, written in C++ and based on  GEANT4 framework~\cite{ref-geant4}.
Used Mokka versions ILD00\_fw and ILD00\_fwp01 offer a realistic description of a forward region of the detector and of the magnetic 
field, which is important in the background studies.
Results presented here were obtained using Marlin~\cite{ref-marlin} which is a C++ analysis and reconstruction framework designed 
for simulated and measured data in ILC-related studies. The concept of Marlin is based on a set of code modules, so called 
processors, that subsequently act on a data stream, read data objects from the stream, perform some sort of action (like. calculation) 
and possibly add new objects to the stream afterwords.

The background studies were based on the expected simulated detector hits from a big pool of bunch crossing (BX) of electron-positron pairs. 
It is clearly seen from the simulation that some of the ILD components are greatly affected by the pair background - vertex detector (VTX) 
and forward detectors, as well as the Time Projection Chamber (TPC). Due to readout times and lack of a time-stamp both the VTX and the TPC 
integrate over a large number 
of background BXs for every physics event. To study effects of the pair background on the reconstruction at ILD we need a good method to
impose background events on real physics events. Now two such methods exist, the detailed description can be found in following sections. 
\subsection{Salt and paper hits}
\begin{wrapfigure}{r}{0.5\columnwidth}
\centerline{\includegraphics[width=.49\columnwidth]{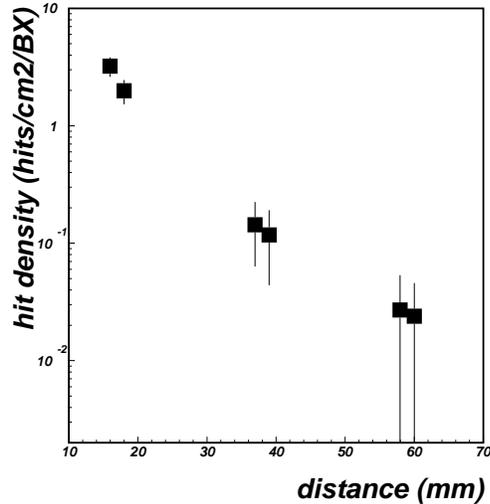}}
\caption{Hit densities for the ILD vertex detector, calculated using GP samples.}\label{hit-den}
\end{wrapfigure}
One way of imposing the pair background is adding background hits to the vertex detector. It is done in Marlin using a VTXNoiseHits 
processor.
Isotropically distributed hits are added to the SimTrackerHits collection after digitizing, according to hit densities. 
Taking into account  estimated readout times for the VTX 83 BXs per event are added for layers 1-2 of the VTX and 333 BXs per event 
for the rest of the layers.
Figure~\ref{hit-den} shows the hit densities evaluated from number of hits in the six vertex detector layers, using 100 BXs of 
Guinea-Pig events simulated using Mokka for nominal beam parameters for the ILD detector~\cite{ref-loi}. 
After adding the hits a full reconstruction and tracking is performed.   

Adding pair background hits to the vertex detector results in a huge amount of additional hits in the VTX and tracks, both in the 
VTX (silicon  tracks) and in the whole detector (full tracks). Table~\ref{tab-numbers} show approximate numbers of the hits and the 
tracks for example physics $b\bar{b}$ events with and without background hits.

Salt and pepper background give a reasonable description for quick studies but do not represent 
realistic enough situation for detailed studies. 
(Not all distribution are isotropic, real tracks are missing and hits should be added in all affected detectors) 
For the fully realistic studies of the effects of the beam background on the reconstruction and tracking the 
overlayed background, described in the following chapter, was used.
\begin{table}
\centerline{\begin{tabular}{|c|c|c|}
\hline
  & no background & with background \\\hline
VTX hits  & ~400 & $10^5$  \\
silicon tracks & ~60 & ~4000 \\
full tracks & ~70 & ~1500\\ 
\hline
\end{tabular}}
\caption{Approximate numbers of hits and tracks for $b\bar{b}$ events with and without background hits.}
\label{tab-numbers}
\end{table}
\subsection{Overlay processor}

For fully realistic studies the OverlayBX Marlin processor was used (version v01-06-fw). It allows overlaying GP 
simulated hits on any physics events. For the detectors with the fast readout (SIT, FTD, SET, ECAL, HCAL, BCAL, LCAL,
LHCAL) 1 BX of GP events is overlayed. In the TPC there is a parameter allowing to set number of overlayed BXs to any number $n$. 
In the vertex detector number of overlayed BX is evaluated from present technology readout times, which corresponds
now to 83 BXs for two inner layers and 333 BXs for the next four layers. For each physics event the overlayed BXs are selected randomly from
a pool of about 2000 GP files (one GP file corresponds to one BX). For the detector with 
longer readout time the overlay processor accounts for time and space shifts for different bunch crossings. 
This method is technically challenging and require big CPU resources but gives the best description of pair background 
overlayed on physics events and was used in the following studies.

\section{Beam background in TPC} 
In this section an example influence of the pair background on the TPC reconstruction and tracking is shown.
Similar studies were done for the vertex detector and are described in details elsewhere~\cite{ref-loi}. 

\begin{figure}
\vspace*{-6cm}
\hspace*{3.5cm}
\centerline{\includegraphics[width=.5\columnwidth]{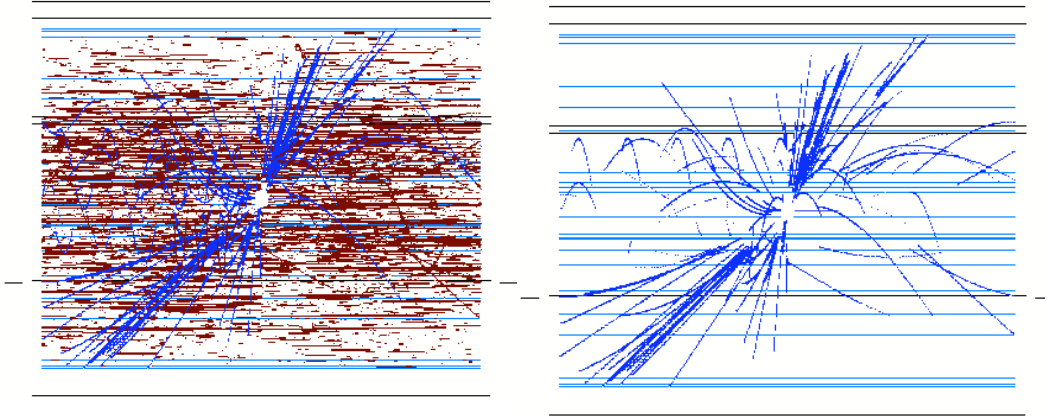}}
\caption{On the left: the $r-z$ view of the TPC hits from a  $t\bar{t}$ event (blue) with 150 BXs of beam background (red) overlayed.
On the right: the same event showing the reconstructed TPC tracks after the micro-curler removal algorithm applied and TPC track
finding done.}
\label{fig-tpc}
\end{figure}

Figure~\ref{fig-tpc}, left, shows the TPC hits for a single $t\bar{t}$ event at $\sqrt{s} = 500$ GeV overlayed with
150 BXs of pair-background hits. On average there are 265,000 background hits in the TPC,
compared to the average number of signal hits of 23100. 
A significant fraction of the background hits in the TPC come from
low energy electrons and positrons from photon conversions. These low energy particles form
small radius helices parallel to the z axis, so called ``micro-curlers'', clearly visible as lines in the $rz$ view. 
Specific pattern recognition software has been written to identify and remove these hits prior
to track reconstruction. 
The cuts remove approximately 99\% of the background hits and only 3\% of hits from the primary interation and 
the majority of
these are from low p$_T$ tracks. Less than 1\% of hits from tracks with $p_T > 1 $~GeV originating
from the $t\bar{t}$ event are removed.
Figure~\ref{fig-tpc}, right, shows that this level of background hits proves no problem for the track-finding pattern 
recognition software and demonstrates the robustness of TPC tracking in the ILC background environment.
\begin{figure}
\hspace*{-3.5cm}
\centerline{\includegraphics[width=.5\columnwidth]{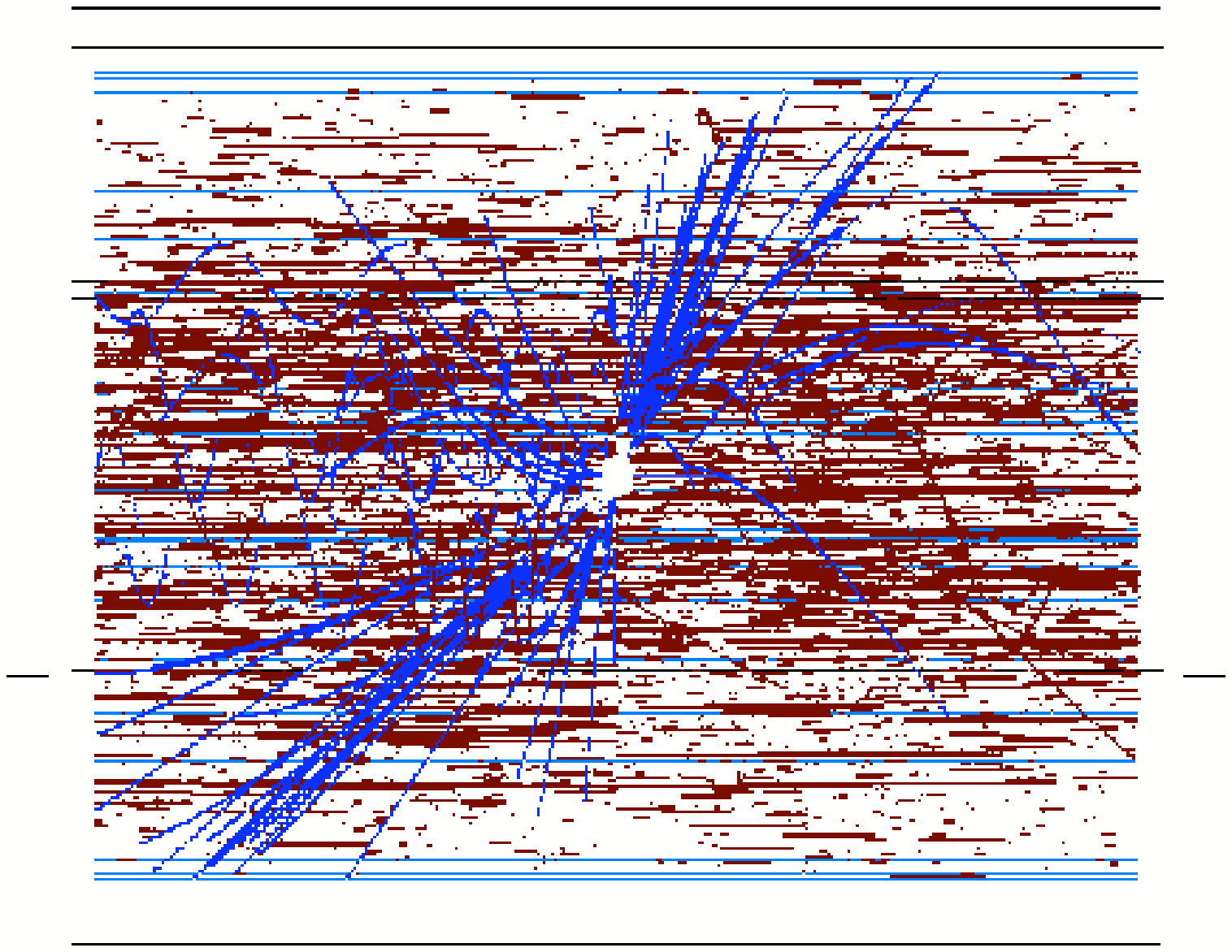}}
\end{figure}
\newpage

\section{Physic analysis with pair background}
\begin{wrapfigure}{r}{0.5\columnwidth}
\centerline{\includegraphics[width=0.45\columnwidth]{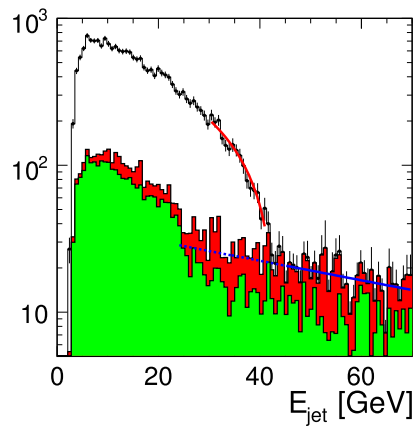}}
\caption{ Tau-jet energy spectrum, black: total, red: total background,
green: SUSY background. Blue line: fit to background.}
\label{fig-stau}
\end{wrapfigure}
To fully simulate the effect of background on a particular physics channel would require
overlaying many BXs of Guinea-Pig background-pair for each detector on each
simulated physics event and would require vast CPU resources. There are other ways to estimate an impact 
of the background on the physics analysis as well but the full simulation should also be considered in future physics studies.
Below an example of a physics analysis with overlayed GP pair-background is presented.~\cite{ref-stau}  

\subsection{Stau in SUSY Sps1a$^{'}$}

For the SUSY SPS1a' parameter set, the process 
$e^+e^- \rightarrow \tilde{\tau}\tilde{\tau} \rightarrow \chi_1 \tau \chi_1 \tau$
gives in the final state missing energy and relatively low energy visible products of tau decays. 
This measurement requires precision tracking of relatively low momentum particles, 
good particle identification, high hermeticity of the detector, and low machine background.

The $\tilde{\tau}$ mass can be extracted from the end-point of the tau-jet energy spectrum, E$_{jet}$, 
shown in Fig.~\ref{fig-stau} and the known $\tilde{\chi}^0_1$ mass.
The statistical uncertainty on the end-point is 0.1 GeV. 
Accounting for the uncertainty on the $\tilde{\chi}^0_1$ mass, $\sigma_{LSP}$, gives a measurement 
precision on $M_{\tilde{\chi}^0_1}$ of $0.1GeV \oplus 1.3\sigma_{LSP}$.

The SUSY study of SPS1a' stau mass, described above  has been repeated with 
a full beam background simulation added to the physics signal. Each physics event has been overlayed 
with 1 BX of $e^+e^-$ pairs.
To remove extra tracks coming from the beam background topological cuts have been used. The cuts require that energy of 
reconstructed particles $E > 0.5$~GeV and each particle has at least one hit in the TPC.
After this additional selection the  DELPHI jet algorithm has been used for tau-jets finding. It is more efficient in a presence 
of extra tracks and clusters from beam background than the standard Durham jet algorithm used in non-background analysis.
The final results of this analysis showed that there is negligible difference in $\tau$ mass extracted
with beam background and without.

\section{Beam background with SB2009}
\begin{wrapfigure}{r}{0.5\columnwidth}
\centerline{\includegraphics[width=0.5\columnwidth]{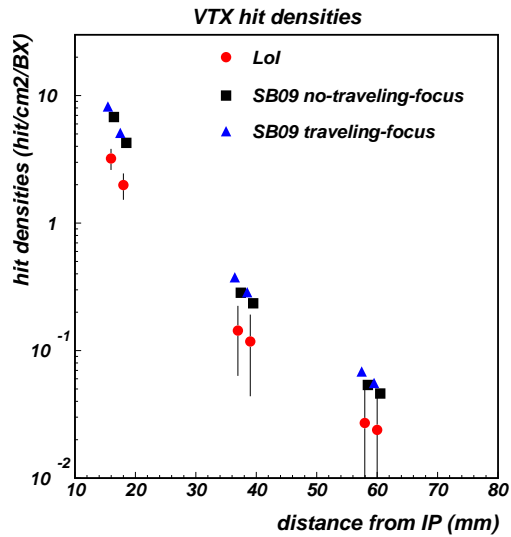}}
\caption{Hit densities for the ILD vertex detector evaluated for the nominal and SB09 beam parameter sets.}
\label{fig-sb09}
\end{wrapfigure}
In September 2009 a new GDE baseline, referred to as Strawman Baseline 2009 or ``SB2009'' was presented and reviewed at 
the Linear Collider Workshop of the Americas, LCWA09.
The ``reduced beam parameter set'', with half the number of bunches
of that planned for the Reference Design, includes changes in the interaction region and beam parameters
and some new techniques are needed to recover full luminosity (for example so called ``traveling focus beam'').
The new set of beam parameters requires reevaluating the background effects on the ILD. 
GP++ samples were generated using
the new reduced beam parameters and simulated using Mokka version ILD00\_fw with GEANT4 range cut of 0.2 mm. 
Hit densities in various detectors were estimated and compared
to the previous results. Two different scenarios were used for the generated GP++ samples: 
with and without traveling focus.
\subsection{Vertex detector}

Figure~\ref{fig-sb09} shows a comparison of the hit densities for the VTX for the nominal and SB09 beam parameters. For the 
SB09 without traveling focus, for 1 BX there is about double amount of hits in the VTX compared to the nominal parameter set.
Using traveling focus adds about 30\% more hits to the no-traveling focus version od SB09. If we take into account that in the
SB09 there is only half as much bunches as in for the nominal set, for the detectors were we integrate hits over some number of
BXs the resulting number of hits will be similar for the nominal set and SB09 without traveling focus and about 15\% higher for 
the traveling focus option. This does not apply to the components with fast readout where we do not integrate over BXs.

Similar analysis for the vertex detector was performed using GP and CAIN samples, simulated with Mokka version 
ILD\_fwp01 and range cut of 0.1 mm.
Number of hits in the VTX for the two studies differs by 25-85\% (depending on the VTX layer) and is caused by the different
GEANT4 range cut. The effect of the range cut on the predicted size of the pair background is under study. 

\subsection{Other detectors}

Similar growth in number of background hits is observed in other ILD detectors: SIT, FTD, SET, TPC, HCAL and ECAL. 
Ratios of SB09 and nominal beam parameters hit numbers range from 1.9 to 2.6 for different detectors for the 
no-traveling-focus option and from 2.5 to 3.6 for the traveling-focus option. The detectors which do not integrate over many BXs
effectively see more pair background that the VTX and the TPC. 
The SB09 setup has as well other consequences, not only growth of the number of hits and these issues are currently under study. Its impact
of the SUSY Sps1a$^{'}$ stau analysis has been presented during this conference and can be found in the proceedings~\cite{ref-stausb09}.

\section{Summary and conclusions}

The ILC faces the novel problem of beam-related backgrounds. It has to be studied in detail and its effects and impact
on the reconstruction and the physics analyzes has to be well understood. 
The pair-background can be generated and simulated 
and there were developed methods of realistic imposing of the background on any physics events. 
For generating the GP background samples 
the normal beam parameter set, as well as SB09 set were considered in the present studies. The SB09 scenario increases 
the pair background by a factor of 2-3 in case of the non-traveling-focus beams and the traveling-focus beams add another
15\% more. 
The studies of beam-induced background help to reduce its impact on the reconstruction and tracking and give us hints 
how to treat this kind of background in the physics analysis. 


\begin{footnotesize}


\end{footnotesize}


\end{document}